\documentstyle[prb,aps,multicol,psfig]{revtex}

\begin{document} 
 
\draft 

\title{Patterned Irradiation of YBa$_2$Cu$_3$O$_{7-x}$ Thin Films}

\author{M.~Basset\cite{byline}$^{,1}$, G.~Jakob$^1$, G.~Wirth$^2$ and 
H.~Adrian$^1$}  

\address{$^1$Institute of Physics, Johannes Gutenberg--Universit\"at, 55099 
Mainz, Germany} 
\address{$^2$Gesellschaft f\"ur Schwerionenforschung, 64291 Darmstadt, 
Germany} 

\date{20 December 2000} 

\maketitle 

\begin{abstract} 
We present a new experiment on YBa$_2$Cu$_3$O$_{7-x}$ thin films using 
spatially resolved heavy ion irradiation. Structures consisting of a 
periodic array of strong and weak pinning channels were created
with the help of metal masks. The channels formed an angle of 
$\pm 45^{\circ}$ with respect to the symmetry axis of the 
photolithographically patterned structures. Investigations of the 
anisotropic transport properties of these structures were performed. We 
found striking resemblance to guided vortex motion as it was observed in 
YBa$_2$Cu$_3$O$_{7-x}$ single crystals containing an array of 
unidirected twin boundaries. The use of two additional test bridges allowed 
to determine in parallel the resistivities of the irradiated and 
unirradiated parts as well as the respective current-voltage characteristics. 
These measurements provided the input parameters for a numerical 
simulation of the potential distribution of the Hall patterning. 
In contrast to the unidirected twin boundaries in our experiment both strong 
and weak pinning regions are spatially extended. The interfaces between 
unirradiated and irradiated regions
therefore form a Bose-glass contact. The experimentally observed magnetic 
field dependence of the transverse voltage vanishes faster than expected 
from the numerical simulation and we interpret this as a hydrodynamical 
interaction between a Bose-glass phase and a vortex liquid.

\end{abstract}



\begin{multicols}{2}
\section{ Introduction}

Heavy ion irradiation has proved to be a powerful method to investigate 
the vortex dynamics of the high temperature superconductors (HTSC). The 
introduced columnar defects (CD) are the most effective pinning centers 
and as a result of heavy ion irradiation the critical current density is 
increased. \cite{civale97} But beside this technological relevance the 
pinning properties of CD have a strong influence on the dynamics of vortices.
Based on the theory of collective pinning \cite{larkin79} and the vortex 
glass model of Fisher, Fisher and Huse (FFH), \cite{fisher91} a Bose-glass 
phase transition is predicted by Nelson and Vinokur for the HTSC compounds 
in the presence of correlated disorder. \cite{nelson92} The underlying 
scaling theory is characterized by universal dynamic and static critical 
exponents $z'$ and $\nu_\perp$ that describe the divergence of the glass 
correlation length $l_\perp (T)$ and the relaxation time $\tau$ at the 
characteristic glass temperature $T_{BG}$. One transport coefficient that 
is related to this glass transition is the shear viscosity $\eta$, controlled 
by the dynamic critical exponent $z'$ via 
$\eta\propto\left| T-T_{BG}\right|^{-z'}$~(Ref.~5).
The divergence of the shear viscosity of the vortex liquid was shown on 
approaching the melting transition temperature of the vortex lattice. 
\cite{pastoriza95,theunissen96}

A different realization of correlated defects is found in twin boundaries 
in YBa$_2$Cu$_3$O$_{7-x}$ (YBCO) single crystals. Their influence on the 
vortex dynamics was extensively studied during the last years in YBCO single 
crystals with unidirected twin boundaries (TB) using electrical 
transport measurements, 
\cite{fleshler93,prodan98,pastoriza99,chabanenko99,Danna00}
ac screening experiments, \cite{jorge00,herbsommer00}
simulations \cite{crabtree96} 
and analytical methods. \cite{sonin93_97,mawatari97_99,shklovskij99}
For the case that the twin boundaries form an angle $\theta$ with the 
direction of an applied external current, the vortices feel an anisotropic 
pinning force. This leads to a guided vortex motion (GVM) that can be 
detected in measuring the even transverse voltage in the presence of an 
external magnetic field. \cite{niessen69} In  literature, two different 
mechanisms are proposed to be responsible for the guided motion of 
vortices in unidirected twins. \cite{crabtree96} In the first, vortices will 
be channeled by the deep pinning potential of the TB and will move in the TB 
(internal motion), whereas the second mechanism assumes the vortex-vortex 
interaction to be the dominating mechanism for a guided vortex motion also 
outside the TB (external motion).

The present work shows a new experiment to address the question of GVM. We 
create strong and weak pinning channels in YBCO thin films using heavy ion 
irradiation through patterned metal masks consisting of an array of fine 
stripes. The effects of homogeneous irradiation are analyzed using irradiated 
and unirradiated reference bridges on the same sample. The transverse voltage 
on a Hall bar structure shows a strong correspondence to that in YBCO 
single crystals with unidirected TB. However, in our experiment the weak 
pinning channels have a macroscopic width in contrast to the very narrow 
channels formed by twin boundaries. This results in a nontrivial current 
density distribution in the samples that is modelled using the longitudinal 
resistivities. Another difference to the planar TBs is that in our case the 
CDs are linelike pinning centers. Therefore a GVM can only result from 
vortex-vortex interaction at the interface of strong and weak pinning 
regions. If the interaction length becomes large the pinned vortices in the 
irradiated channels should induce a freezing of free vortices in the 
unirradiated channels. At this freezing transition the GVM vanishes.
Thus, the arrangement of strong and weak pinning regions represents a 
Bose-glass contact. A Bose-glass contact was introduced by Marchetti 
{\em et al.} \cite{marchetti99} in the definiton of a new generation of 
experiments in order to study the Bose-glass transition in a hydrodynamical 
context. \cite{marchetti90} The voltage current characteristics of both test 
bridges was investigated in order to study the correlations between GVM 
effect and the Bose-glass transition.

\section{ Sample Preparation}

YBCO thin films have been prepared using the dc sputtering technique. 
\cite{trosa90} The characterization with X-ray diffraction and scanning 
electron microscopy shows epitaxial growth and homogeneous surfaces of the 
films. The samples were patterned with standard photolithography and wet 
chemical etching into a conventional Hall structure 
$\rm{(2\;mm \times 7\;mm)}$ and two identical test bridges 
$\rm{(200\;\mu m\times 2000\;\mu m)}$.The irradiation of the samples was 
performed at the UNILAC of the \textit{Gesellschaft f\"ur 
Schwer\-ionenforschung}, GSI. The samples were irradiated with $0.75$~GeV 
Pb-ions and $1.14$~GeV U-ions in such a way that CD were arranged in a 
periodic array of strong (irradiated) and weak (unirradiated) pinning 
regions. Differently patterned  $\rm{0.5\;mm}$ thick Ni-masks were used in 
order to stop the swift ions and to reproduce the irradiation pattern on the 
thin film structure. 

\begin{figure}[t]
\centerline{\psfig{file=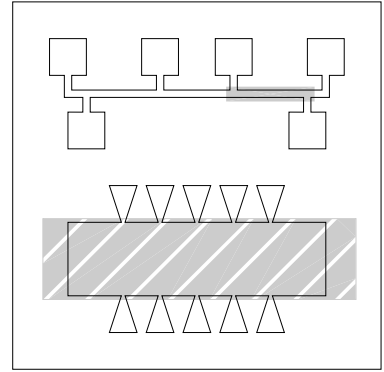,width=0.6\columnwidth}}
\vspace{1.5em} 
\caption{Sample and irradiation geometry of the 1:7 ratio structure. The 
	upper part shows the two test bridges. Grey regions are irradiated.}
\label{structure}
\end{figure}

\begin{figure}[t]
\psfig{file=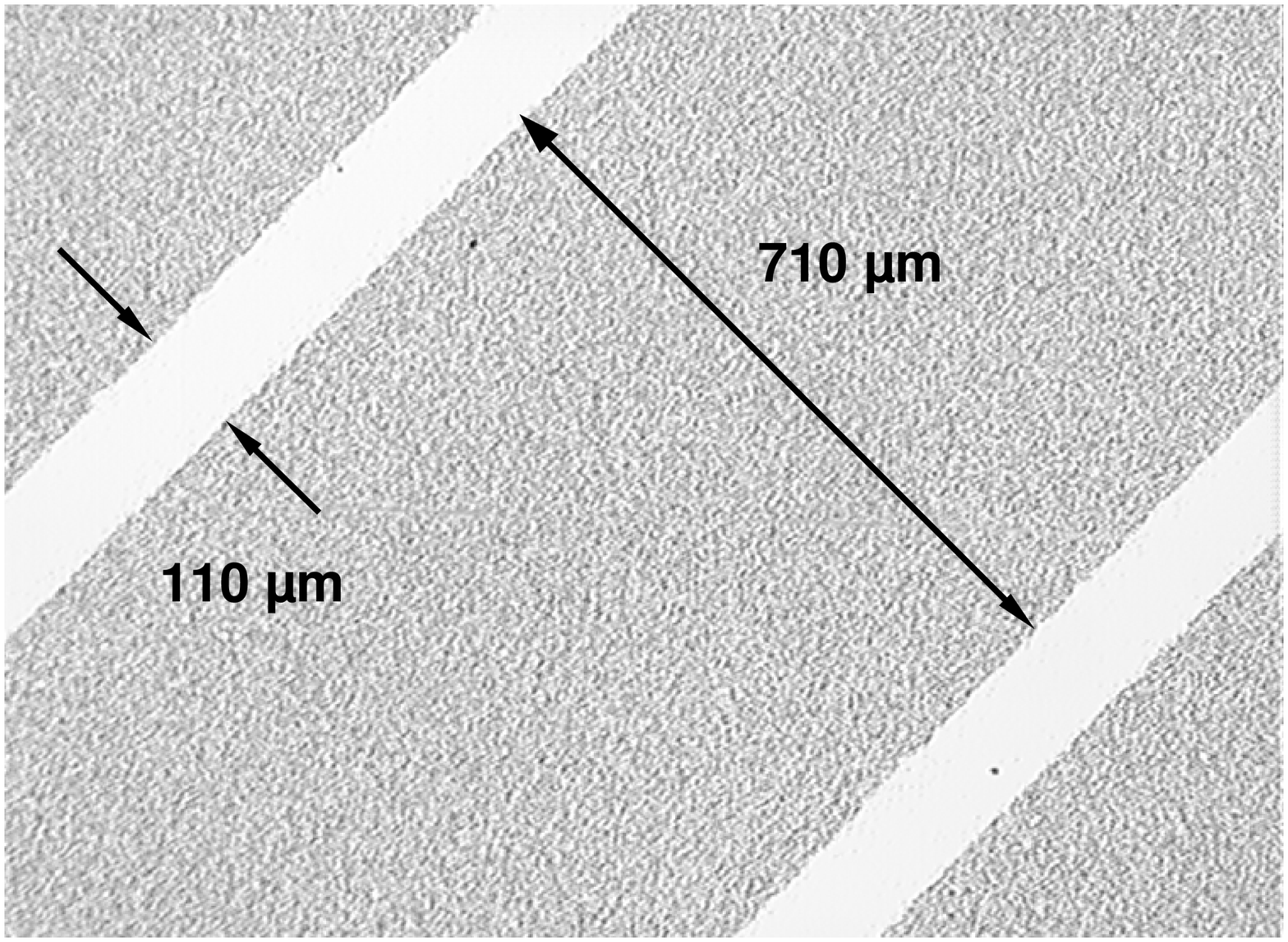,width=0.9\columnwidth} 
\vspace{0.5em} 
\caption{Image of an irradiated glass substrate that was etched in 18\% HF 
	for ${\rm 20\;s}$. For the structure with the 1:7 ratio that is shown 
	here, the channel width is ${\rm 110\;\mu m}$ for the unirradiated 
	and ${\rm 710\;\mu m}$ for the irradiated regions.}
\label{glass} 
\end{figure}

Two arrangements were used in this experiment which differ only in the width 
of the irradiated channels that is ${\rm 700\;\mu m}$ (1:7 ratio) and 
${\rm 400\;\mu m}$ (1:4 ratio) respectively, whereas the width of the weak 
pinning channel is ${\rm 100\;\mu m}$ for each case.
The angle $\theta$ between the edges of the Hall structure and the direction 
of the weak pinning channels is $\pm 45^\circ$. In addition, one test bridge 
was irradiated entirely with the same ion dose 
($B_\phi=1.0\;{\rm T}\;\widehat{=}\;5\cdot 10^{10}\;{\rm Ions}/{\rm cm}^2$) 
as the Hall structure. Sample and irradiation geometry are sketched in 
Fig.~\ref{structure} for the Hall structure with the 1:7 ratio. All samples 
show sharp resistive zero field transitions $(\Delta T<1 \;{\rm K}$) for 
both, the irradiated and unirradiated test bridges with critical temperatures 
of $T_c\approx 91\;{\rm K}$. The efficiency of the metal masks in the 
patterned irradiation was analyzed using glass substrates that were 
irradiated through the masks with a dose of approximately 
$3\cdot 10^{6}\;\rm{Ions/cm^2}$. To visualize the defects, the glass 
substrates were etched with $\rm{18\%\;HF}$ for $\rm{20\;s}$. A result of 
this procedure is shown in Fig.~\ref{glass} for the 1:7 ratio. The width of 
the weak pinning channels is $\rm{110\;\mu m}$, the periodic length of the 
irradiation pattern is $\rm{820\;\mu m}$. Higher image resolutions show that 
the defect concentration has a step-like behavior at the interfaces between 
strong and weak pinning. Less than $\rm{2\%}$ of the ions are scattered in a 
$\rm{5\;\mu m}$ wide region of the unirradiated channels in both experimental 
arrangements.

\section{ Experiment} 

Measurements of the voltage drop across different contacts of the Hall 
structure were performed for temperatures from ${\rm 300\;K}$ to 
${\rm 70\;K}$ and magnetic fields up to $12\;{\rm T}$ at a constant current 
density of $J=125\;{\rm A/cm^2}$. The various contacts allow to measure the 
voltage drop in different directions with respect to the direction of the 
weak pinning channels and the external current. In addition, the two parts of 
the test bridge allow to determine in parallel the voltage drop over a 
completely irradiated and unirradiated part of the sample for all parameters 
used in the measurement. These results represent the input values for scaled 
resistor network representations of the periodically irradiated Hall 
structures with the different arrangements. Based on Kirchhoff's laws, 
it is then possible to calculate the potential distribution at the borders of 
the Hall structures and to compare the calculated with the measured results. 
Another advantage of this experimental set is the possibility to measure the 
current-voltage characteristics of both, the unirradiated and irradiated part 
of the film. Based on this information one can carry out a glass analysis in 
order to determine the characteristic fields and temperatures of the vortex 
dynamics for both test bridges.

\section{ Results}

\begin{figure}[t]
\psfig{file=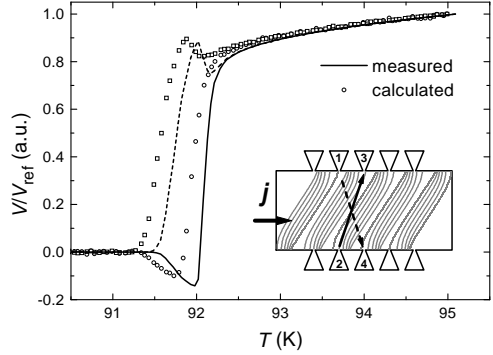,width=0.9\columnwidth} 
\vspace{0.5em} 
\caption{Voltage drop between two contact pairs of different direction with 
	respect to the applied current density for the structure with the 1:7 
	ratio. The solid line (calculated data) corresponds to the open 
	circles (measured results), the dashed line to open squares. 
	Measured and calculated data are normalized to its corresponding 
	value at $T=95\;{\rm K}$. The position of the contacts is sketched 
	schematically in the inset. In this sample $T_c$ differs by 
	$0.14\;{\rm K}$ between test bridges and Hall structure, visible as a 
	shift in the local extrema between calculated and measured curves. 
	The inset shows the measurement geometry as well as the equipotential 
	lines for $T=91.88\;{\rm K}$.}
\label{nullfeld} 
\end{figure}

The resistive transitions of the irradiated Hall structure have been measured 
for different directions of the voltage drop with respect to the external 
current direction in zero field. In an unirradiated sample transverse 
voltages are negligible in zero field and one should observe a monotonously 
vanishing voltage signal along all contact configurations containing a 
longitudinal component. Measuring the configurations connecting contacts 1 
and 3 ($V_{13}$) (see Fig.~\ref{nullfeld}) should yield identical results 
as testing $V_{24}$, $V_{14}$, or $V_{23}$, respectively. In the patterned 
irradiation geometry the symmetry of the contact configuration is broken. The 
resistive transition measured using $V_{14}$ shows a maximum
at $91.88\;\rm{K}$ and using $V_{23}$ one even observes a sign reversal. 
The network calculation reproduces these unusual features for both contact 
configurations. Figure~\ref{nullfeld} shows the temperature dependence of 
$V_{14}$ and $V_{23}$ together with the corresponding calculations. The 
existence of a sign reversal for $V_{23}$ is obvious from the potential 
distibution sketched in the inset. It shows the equipotential lines across 
the 1:7 ratio structure resulting from the network calculation at 
$T=91.88\;{\rm K}$. By measuring the two indicated directions, the number of 
lines that are crossed differs due to the arrangement of the strong and weak 
pinning channels. Following the dashed line, i.e. $V_{14}$, the equipotential 
lines are only crossed in the positive current direction, whereas this is 
not the case for the contacts of the solid line ($V_{23}$). 

If the voltage drop for fixed temperatures is measured during magnetic field 
sweeps ($-12\;{\rm T}\;<\mu_0H<12\;{\rm T}$), the consistency between 
calculation and measurement holds for the longitudinal voltages as well as 
for the transverse case. The network calculation does not include the 
asymmetric part of the transverse voltages $V^t_{\rm odd}$ that can be 
measured for inversion of the magnetic field due to the Hall effect. 
\cite{chien91} However, here we are interested in the GVM due to anisotropic 
pinning forces. This shows up as a transverse voltage, symmetric with respect 
to the field direction \cite{niessen69}

		\begin{equation}
		    \label{even}
		    V^t_{\rm even}=\frac{V_{12}^t(+H)+V_{12}^t(-H)}{2}.
		\end{equation}

\begin{figure}[t]
\psfig{file=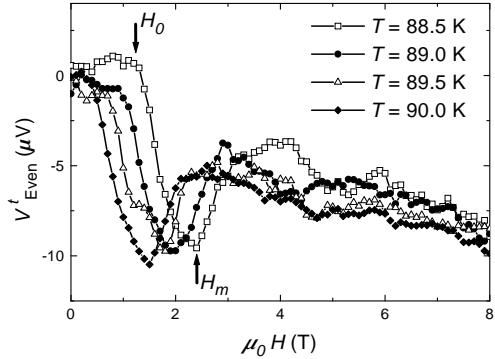,width=0.9\columnwidth} 
\vspace{0.5em} 
\caption{Even transverse voltage for temperatures below $T_c$ in the case of 
	the 1:7 ratio structure. The curve is characteristic for the GVM and 
	was also observed for YBCO single crystals with unidirected TB.
	The characteristic fields are considered to be $\mu_0H_m$ where 
	$V^t_{\rm even}$ reaches its maximum and $\mu_0H_0$ where it 
	vanishes.}
\label{guiding} 
\end{figure}

\begin{figure}[t]
\psfig{file=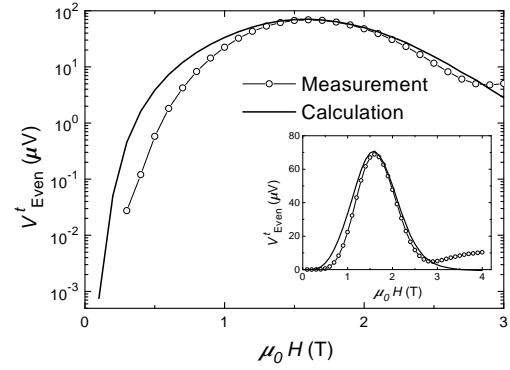,width=0.9\columnwidth} 
\vspace{0.5em} 
\caption{Even transverse voltage of the Hall-structure with 1:4 ratio at a 
	temperature of $88\;{\rm K}$. Near the transition to 
	$V^{t}_{\rm even}=0$ calculated and measured values differ by one 
	order of magnitude. The inset shows a linear plot of the transition. 
	The deviation for high magnetic fields is discussed in the text.}
\label{guiding1_4} 
\end{figure}

This procedure eliminates the asymmetric Hall contributions.
The magnetic field dependence of $V^t_{\rm even}$ at different temperatures 
is shown in Fig.~\ref{guiding} for the 1:7 ratio structure and in 
Fig.~\ref{guiding1_4} for the 1:4 ratio structure, respectively. Both samples 
show the same behavior, except for the fact that the angle $\theta$ 
describing the direction of the weak pinning channels is $+45^\circ$ in one 
and $-45^\circ$ in the other case. This results in negative and positive 
values for $V^t_{\rm even}$ as can be directly seen from the calculation of 
the potential distribution (e.g. inset of Fig.~\ref{nullfeld}) for the case 
of $\theta=+45^\circ$. For magnetic fields $\mu_0H>3\;{\rm T}$, the absolute 
value decreases with decreasing field. For smaller fields, an increase of 
$\left|V^{t}_{\rm even}\right|$ which has a maximum at $\mu_0H_m$ followed by 
a sharp transition to $V^{t}_{\rm even}\!=\!0$ at $\mu_0H_0$ can be observed. 
The characteristic field $\mu_0H_m$ increases and 
$\left|V^{t}_{\rm even}(\mu_0H_m)\right|$ decreases with decreasing 
temperature. This behavior is exactly the same as observed in YBCO single 
crystals with unidirected twin boundaries. \cite{chabanenko99} Thus, these 
results are a first indication for GVM in periodically irradiated YBCO films.
Compared to YBCO single crystals one should note that the length scale of 
both experiments is very different. The usual periodic length in TB 
experiments is about $\rm{1\;\mu m~to~10\;\mu m}$, 
\cite{pastoriza99,chabanenko99} whereas the experiment here was performed on 
a length scale that is two orders of magnitude larger. In spite of this huge 
difference the results obtained in both cases are comparable. At high 
field values in our experiment the resistivity of irradiated and unirradiated 
regions of the film is close to the normal state resistivity, which 
is nearly the same for both regions and therefore no transverse voltage 
should occur. For single crystals in the normal state the twin boundaries 
only marginally influence the total resistivity and accordingly the 
transverse voltage vanishes. Experimentally small values of $V^t_{\rm even}$ 
are observed in both cases which are attributed to a small longitudinal 
contribution. For temperatures below $T_c$ and low magnetic fields again no 
transverse voltage is observed in both experiments. Clearly in this 
temperature and field range the vortex system becomes rigid and the 
longitudinal resistivity vanishes in the whole sample enforcing zero 
transverse voltage.

Therefore, we focus in the following on the intermediate field range where a 
pronounced maximum of $V^{t}_{\rm even}$ is observed. Our calculation 
reproduces quantitatively the maximum in the magnetic field dependence of 
$V^{t}_{\rm even}$ as shown in Fig.~\ref{guiding1_4}. The existence of this 
maximum was already verified in an analytical approach by Shklovskij 
{\em et al.} \cite{shklovskij99} They used an effective medium approach to 
calculate the averaged resistivity of YBCO crystals containing dense arrays 
of unidirected twin planes. However, in our case we only have a small number 
of weak pinning channels and an effective medium theory is not appropriate.
A second point to mention is that possible nonlinearities of the 
current-voltage characteristics (CVCs) have to be taken into account. This is 
done by using identical current densities for the reference bridges and the 
Hall structures. In this case the continuity condition for the current 
eliminates nonlinear effects in first approximation. Compared to the 
calculated values for $V^t_{\rm even}$ the experimentally determined values 
show deviations for high and low fields. At high fields, the influence of the 
limited lateral resolution of the photolithographically patterned voltage 
probes leads to a residual longitudinal voltage. This can be included in the 
network calculations and one can show that this contribution is negligible at 
fields below $\mu_0H_m$. More interesting are the deviations for $H<H_0$ 
where the transition to $V^{t}_{\rm even}=0$ appears. The logarithmic 
illustration of Fig.~\ref{guiding1_4} indicates that the measured value is 
nearly one order of magnitude smaller than the corresponding calculated value 
expected from the longitudinal resistivities of the reference bridges. We 
believe that this discrepancy is due to the interaction between pinned flux 
lines in the irradiated channels and vortices in the weak pinning channels as 
will be discussed below.

\begin{figure}[t]
\psfig{file=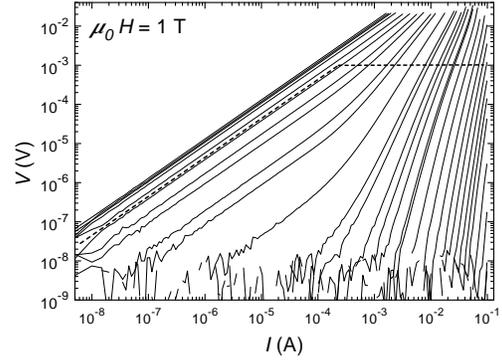,width=0.9\columnwidth} 
\vspace{0.5em} 
\caption{Current-voltage characteristics of the irradiated test bridge for a 
	magnetic field $\mu_0H=1\;{\rm T}$ and temperatures between 
	$91\;{\rm K}$ and $80\;{\rm K}$. The dashed lines border the region 
	that was used for the Bose-glass scaling shown in Fig.~\ref{Bose}.}
\label{Uvi} 
\end{figure}

A further interpretation of these results requires a detailed analysis of the 
different vortex states in the irradiated Hall structure. A straightforward 
access to this problem is the analysis of the current-voltage characteristics 
of the test bridges. Figure~\ref{Uvi} shows such a set of curves for the 
irradiated part of the test bridge and a magnetic field of 
$\mu_0H=1\;{\rm T}$. The Bose-glass temperature $T_{BG}$ of the irradiated 
and the vortex glass temperature $T_{VG}$ of the unirradiated test bridge 
were determined using scaling analysis. \cite{koch89,nelson93}

\begin{figure}[t]
\psfig{file=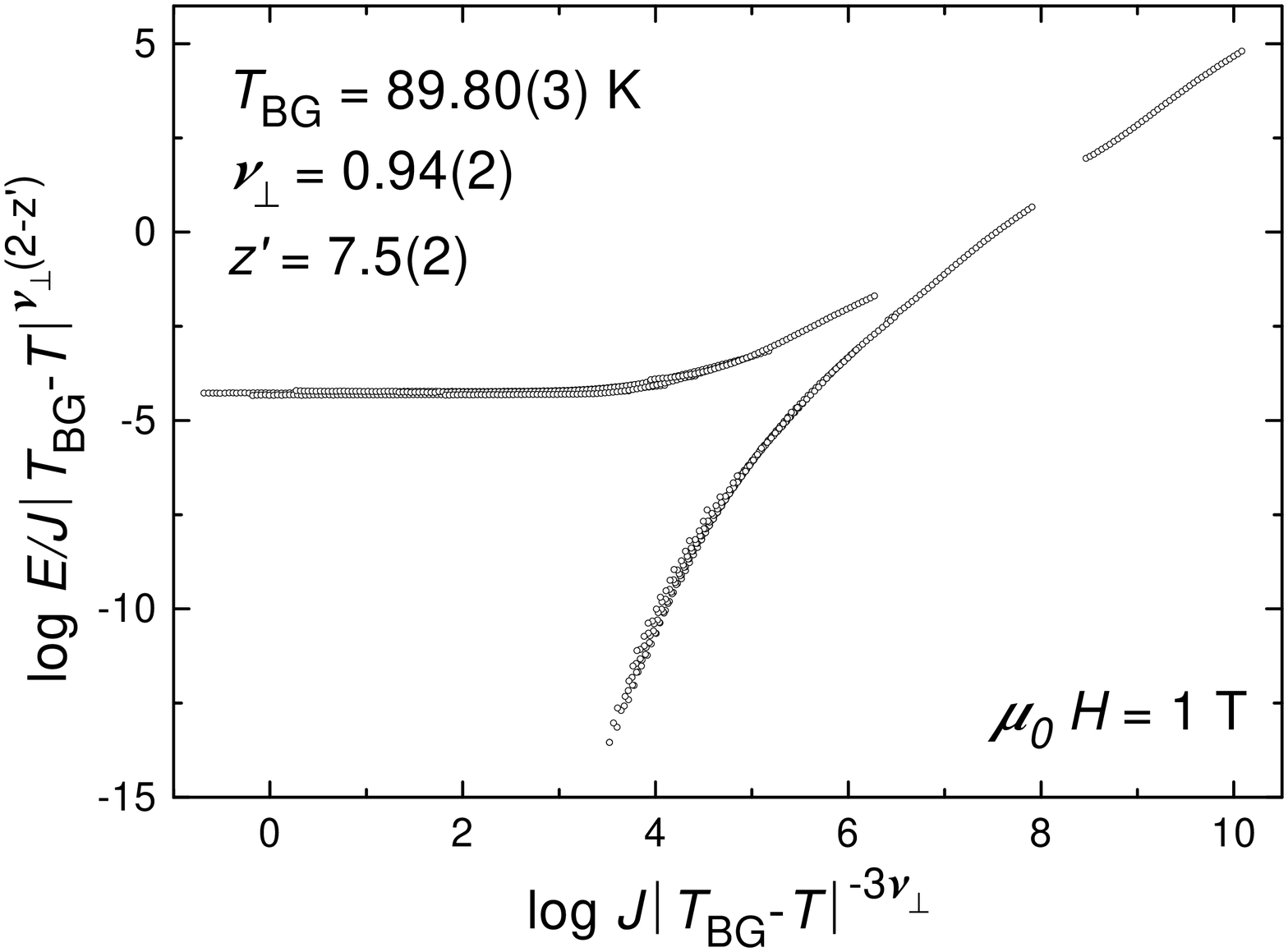,width=0.9\columnwidth} 
\vspace{0.5em} 
\caption{Bose-glass scaling of the $I-V$ curves of Fig.~\ref{Uvi} in 
	accordance with Eq.~\ref{scaling}. Good scaling is achieved with 
	$T_{\rm BG}=89.8(3)\;{\rm K}$, $\nu_\perp=0.94(2)$ and $z'=7.5(2)$ 
	and the determined critical exponents are in agreement with 
	theoretical predictions. \cite{blatter94}}
\label{Bose} 
\end{figure}

While vortex-glass and Bose-glass represent two
distinct physical phenomena the scaling relations are mathematically 
equivalent but differ in the critical exponents.
In both cases the $I-V$ curve that separates the glassy from the fluid regime 
gives a power law dependence over the whole current range. \cite{blatter94}
For the Bose-glass case this relation between electric field $E$ and current 
density $J$ is given by
		\begin{equation}
		    \label{slopeTG}
		    E\propto J^{(z'+1)/3},
		\end{equation}
where $z'$ is the dynamic exponent of the Bose-glass transition. 
\cite{nelson92}\\
Starting with a scaling ansatz \cite{nelson93}

		\begin{equation}
		    \label{scaling}
		    E\propto l_\perp^{-(z'+1)}{\cal E}_\pm (Jl_\perp l_\|)
		\end{equation}
and supposing that the parallel glass correlation length $l_\|$ is 
proportional to $l_\perp^2$ (threedimensional case), one can plot all the 
$I-V$ curves lying below the dashed lines in Fig. \ref{Uvi}. Such a scaling 
is sketched in Fig. \ref{Bose} for the irradiated test bridge at 
$\mu_0H=1\,\rm{T}$. For $z'=7.5(2)$ and $\nu'_\perp=0.94(2)$, all $I\!-\!V$ 
curves in question collapse onto two branches that are separated by 
$T_{BG}=89.8(3)\,\rm{K}$.

\begin{figure}[t]
\psfig{file=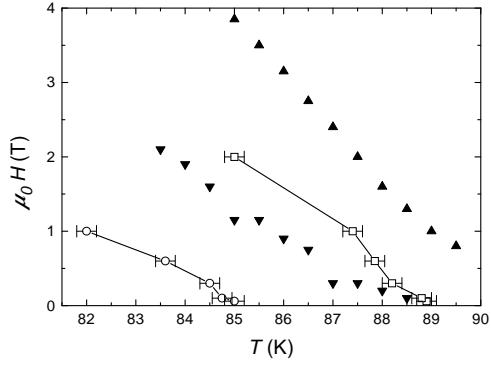,width=0.9\columnwidth} 
\vspace{0.5em} 
\caption{Temperature dependent characteristic fields of the irradiated (open 
	squares, $B_{\rm BG}(T)$) and unirradiated (open circles, 
	$B_{\rm VG}(T)$) test bridge. In addition, $B_0(T)$ revealing from 
	the magnetoresistive measurements of $V^t_{\rm even}$ for the 1:4 
	ratio structure is plotted (filled down triangles) together with the 
	$B_{\rm max}(T)$ dependence (filled up triangles). The position and 
	the behavior of the $B_0(T)$ curve in the phase diagram corresponds 
	to predictions that were made for a Bose-glass contact.}
\label{BvTg} 
\end{figure}

Analyzing the Bose-glass transition for different magnetic fields, one 
obtains the irreversibility line defined by $T_{BG}(B)$. \cite {nelson93} 
The temperature dependence of the characteristic field $B_{BG}(T)$ is shown 
in Fig.~\ref{BvTg} (open squares) together with the $B_{VG}(T)$-behavior of 
the unirradiated test bridge (open circles). It is in qualitative accordance 
with $H(T)$ diagrams obtained in experiments that reveal the influence of 
correlated defects on the vortex system. \cite{pastoriza95,herbsommer00} 
In addition, the characteristic points of $V^t_{\rm even}(B)$ for different 
temperatures shown in Fig.~\ref{guiding1_4} are plotted. The magnetic fields 
$B_{\rm max}$ at which $V^t_{\rm even}$ reaches its maximum (filled up triangles) 
and $B_0$ where $V^t_{\rm even}$ disappears (filled down triangles) are two 
important features. 

The first information one can extract from this diagram is that the maximum 
of $V^t_{\rm even}$ appears for fields and temperatures above the 
irreversibility line of the irradiated test bridge. A comparison between the 
1:4-ratio and the 1:7-ratio structure shows that the temperature dependence 
of $B_{\rm max}$ is proportional to $T/T^{\rm irr}_c$, where $T^{\rm irr}_c$ 
is the critical temperature of the irradiated test bridge in zero field. 
Thus, the peaks in the field dependence of $V^t_{\rm even}$ appear in the 
flux flow regime as shown in Fig.~\ref{guiding} and Fig.~\ref{guiding1_4}.

The analysis of the temperature and magnetic field behavior of 
$V^t_{\rm even}=0$ is the second point to be discussed in this context. In 
the phase diagram the obtained data points for $B_0(T)$ can be found between 
the irreversibility lines of irradiated and unirradiated test bridges. 
However, the position of the ($B_0(T)$)--line with respect to the test bridge 
results depends on the ratio between the widths of both, strong and weak 
pinning channels. For the 1:7 ratio structure the irradiated channels are 
dominant and the ($B_0(T)$)--line coincides with the irreversibility line of 
the irradiated test bridge. The situation is different for the Hall structure 
with the 1:4 ratio. In this case, the ($B_0(T)$)--line is shifted towards 
lower fields and lower temperatures.

To explain these results, a hydrodynamic approach at the vortex Bose-glass 
transition as proposed by Marchetti and Nelson \cite{marchetti99,marchetti00} 
is considered. This approach is based on two heavily irradiated Bose-glass 
contacts that sandwich a weak pinning channel. Viewing on the potential 
distribution sketched in the inset of Fig.~\ref{nullfeld} shows that our 
experiment represents this scenario.
The different glass temperatures lead to a vortex motion in the weak pinning 
channel whereas the vortices in the strong pinning regions are trapped, 
acting like a barrier for the free vortices. The influence of the Bose-glass 
contacts can be described by a viscous length $\delta$, depending on the the 
flux liquid viscosity. \cite{marchetti90} A further analysis shows that 
$\delta$ is the Bose-glass correlation length $l_\perp$~(Ref.~4). Therefore, 
the borders of a weak pinning channel can be understood as a Bose-glass 
contact and we interpret the results of Fig.~\ref{guiding} and 
Fig.~\ref{guiding1_4} in this sense. The vanishing transverse resistivity at 
$B_0(T)$ indicates the transition into a glass phase in the weak pinning 
channels at fields well above $B_{VG}$. This is underlined by the comparison 
between the measured and calculated magnetic field dependence of 
$V^t_{\rm even}$. The transition to $V^t_{\rm even}=0$ is measured for higher 
fields than the calculated transition, indicating that the divergence of the 
shear viscosity $\eta$ depends on whether strong and weak pinning regions are 
independent or, as in the case of the Hall structures and Bose-glass 
contacts, depends on the interaction between vortices in both regions. The 
length scale set by this interaction can be macroscopic and is described by 
the viscous length $\delta$. 

This result is consistent with other experiments carried out on this field of 
research. The diverging correlation length of the vortex system near the 
melting transition was already predicted by Nelson and Halperin. 
\cite{nelson79} For ${\rm Bi_2Sr_2CaCu_2O_8}$ single crystals being 
irradiated similarly to the samples used in this experiment, Pastoriza 
{\em et al.} \cite{pastoriza95} pointed out that the characteristic features 
of the melting transition can be observed when the correlation length is 
comparable to the dimensions of the width of the unirradiated channel 
(in this case $\approx10\;\mu {\rm m}$). 
Experiments in the Corbino disk geometry give an estimation of the order of 
magnitude of the dynamic correlations between vortices in YBCO single 
crystals. L\'opez and co-workers \cite{lopez99} report of a correlation 
length with macroscopic dimensions. The voltage probes they used to obtain 
these results have distances of $60\;\mu {\rm m}$ which is in good agreement 
to the dimensions of the unirradiated channels in the experiment described 
above.

\section{Conclusion}
In summary, we have investigated the vortex dynamics in periodically 
irradiated YBCO thin films for $100\;\mu {\rm m}$ wide unirradiated channels 
forming an angle of $\pm 45^\circ$ with the direction of an external current.
Compared to experiments on single crystals with unidirected TB this method is 
not limited on YBCO and can also be applied to other HTSC thin films and 
single crystals. Another important advantage of the described method is that 
there are no restrictions to the geometry of the heavy ion irradiation 
pattern. The even transverse voltage gives clear indications for a guided 
motion of the vortices in such systems as it was observed in recent 
experiments on YBCO single crystals with unidirected twin boundaries.

With the aid of an irradiated and an unirradiated test bridge it was possible 
to gain input parameters for a numerical simulation of the potential 
distribution and to analyze the current-voltage characteristics of both the 
strong and weak pinning parts, respectively. Based on these results, the 
existence of a Bose-glass contact in the periodically irradiated samples was 
assumed and the obtained $B-T$ phase diagram was interpreted in a 
hydrodynamical approach near the Bose-glass transition.

\acknowledgments

The authors thank H.~Lott for preparing the irradiation masks as well as 
E.~J\"ager and E.~Schimpf for experimental assistance during irradiation. 
V.A.~Shklovskij and A.K.~Soroka are acknowledged for valuable comments and 
discussions. 

This work was supported by the GSI under contract No.~MZADRM.

\bibliographystyle{prsty}

\begin{references} 
\bibitem[*]{byline}Electronic address: basset@mail.uni-mainz.de

\bibitem{civale97}
L. Civale, Supercond. Sci. Technol. {\bf 10},  A11  (1997).

\bibitem{larkin79}
A.~I. Larkin and Y.~M. Ovchinnikov, J. Low Temp. Phys. {\bf 34},  409  (1979).

\bibitem{fisher91}
D.~S. Fisher, M.~P.~A. Fisher, and D.~A. Huse, Phys. Rev. B {\bf 43},  130
  (1991).

\bibitem{nelson92}
D.~R. Nelson and V.~M. Vinokur, Phys. Rev. Lett. {\bf 68},  2398  (1992).

\bibitem{marchetti99}
M.~C. Marchetti and D.~R. Nelson, Phys. Rev. B {\bf 59},  13624  (1999).

\bibitem{pastoriza95}
H. Pastoriza and P.~H. Kes, Phys. Rev. Lett. {\bf 75},  3525  (1995).

\bibitem{theunissen96}
M.~H. Theunissen, E. {Van der Drift}, and P.~H. Kes, Phys. Rev. Lett. 
{\bf 77}, 159  (1996).

\bibitem{fleshler93}
S. Fleshler, W.~K. Kwok, U. Welp, V.~M. Vinokur, M.~K. Smith, J. Downey, and
  G.~W. Crabtree, Phys. Rev. B {\bf 47},  14448  (1993).

\bibitem{prodan98}
A.~A. Prodan, V.~A. Shklovskij, V.~V. Chabanenko, A.~V. Bondarenko, M.~A.
  Obolenskii, H. Szymczak, and S. Piechota, Physica C {\bf 302},  271  (1998).

\bibitem{pastoriza99}
H. Pastoriza, S. Candia, and G. Nieva, Phys. Rev. Lett. {\bf 83},  1026
  (1999).

\bibitem{chabanenko99}
V.~V. Chabanenko, A.~A. Prodan, V.~A. Shklovskij, A.~V. Bondarenko, M.~A.
  Obolenskii, H. Szymczak, and S. Piechota, Physica C {\bf 314},  133  (1999).

\bibitem{Danna00}
G. D'Anna, V. Berseth, L. Forr\'o, A. Erb, and E. Walker, Phys. Rev. B {\bf
  61},  4215  (2000).

\bibitem{jorge00}
G.~A. Jorge and E. {Rodr\'{\i}guez}, Phys. Rev. B {\bf 61},  103  (2000).

\bibitem{herbsommer00}
J.~A. Herbsommer, G. Nieva, and L. Luzuriaga, Phys. Rev. B {\bf 61},  11745
  (2000).

\bibitem{crabtree96}
G.~W. Crabtree, G.~K. Leaf, H.~G. Kaper, V.~M. Vinokur, A.~E. Koshelev, D.~W.
  Braun, D.~M. Levine, W.~K. Kwok, and J.~A. Fendrich, Physica C {\bf 263},
  401  (1996).

\bibitem{sonin93_97}
E.~B. Sonin, Phys. Rev. B {\bf 48},  10487  (1993); {\bf 55},  485  (1997).

\bibitem{mawatari97_99}
Y. Mawatari, Phys. Rev. B {\bf 56},  3433  (1997); {\bf 59},  12033  (1999).

\bibitem{shklovskij99}
V.~A. Shklovskij, A.~K. Soroka, and A.~A. Soroka, JETP Lett. {\bf
  89},  1138  (1999).

\bibitem{niessen69}
A.~K. Niessen and C.~H. Weijsenfeld, J. Appl. Phys. {\bf 40},  384  (1969).

\bibitem{marchetti90}
M.~C. Marchetti and D.~R. Nelson, Phys. Rev. B {\bf 42},  9938  (1990).

\bibitem{trosa90}
C. Tom\'e-Rosa, G. Jakob, M. Maul, A. Walkenhorst, M. Schmitt, P. Wagner, P.
  Przyslupski, and H. Adrian, Physica C {\bf 171},  231  (1990).

\bibitem{chien91}
T.~R. Chien, T.~W. Jing, N.~P. Ong, and Z.~Z. Wang, Phys. Rev. Lett. {\bf 66},
  3075  (1991).

\bibitem{koch89}
R.~H. Koch, V. Foglietti, W.~J. Gallagher, G. Koren, A. Gupta, and M.~P.~A.
  Fisher, Phys. Rev. Lett. {\bf 63},  1511  (1989).

\bibitem{nelson93}
D.~R. Nelson and V.~M. Vinokur, Phys. Rev. B {\bf 48},  13060  (1993).

\bibitem{blatter94}
G. Blatter, M.~V. Feigel'man, V.~B. Geshkenbein, A.~I. Larkin, and V.~M.
  Vinokur, Rev. Mod. Phys. {\bf 66},  1125  (1994).

\bibitem{marchetti00}
M.~C. Marchetti and D.~R. Nelson, Physica C {\bf 330},  105  (2000).

\bibitem{nelson79}
D.~R. Nelson and B.~I. Halperin, Phys. Rev. B {\bf 19},  2457  (1979).

\bibitem{lopez99}
D. L\'opez, W.~K. Kwok, H. Safar, R.~J. Olsson, A.~M. Petrean, L. Paulius, and
  G.~W. Crabtree, Phys. Rev. Lett. {\bf 82},  1277  (1999).

\end{references}

\end{multicols}
\end{document}